# Bulk and Surface Tunneling Hydrogen Defects in Alumina

Aaron M. Holder[a,b], Kevin D. Osborn[c], C. J. Lobb[d], and Charles B. Musgrave[a,b]

[a] *Department of Chemistry and Biochemistry, University of Colorado, Boulder, Colorado 80309*
[b] *Department of Chemical and Biological Engineering, University of Colorado, Boulder, Colorado 80309*
[c] *Laboratory for Physical Sciences, College Park, Maryland 20740*
[d] *Department of Physics, CNAM and JQI, University of Maryland, College Park, Maryland 20742*

We perform *ab initio* calculations of hydrogen-based tunneling defects in alumina to identify deleterious two-level systems (TLS) in superconducting qubits. The defects analyzed include bulk hydrogenated Al vacancies, bulk hydrogen interstitial defects, and a surface OH rotor. The formation energies of the defects are first computed for an Al- and O-rich environment to give the likelihood of defect occurrence during growth. The potential energy surfaces are then computed and the corresponding dipole moments are evaluated to determine the coupling of the defects to an electric field. Finally, the tunneling energy is computed for the hydrogen defect and the analogous deuterium defect, providing an estimate of the TLS energy and the corresponding frequency for photon absorption. We predict that hydrogenated cation vacancy defects will form a significant density of GHz-frequency TLSs in alumina.

Superconducting qubits and resonators allow simple quantum information algorithms to be performed in integrated circuits [1,2]. Unfortunately, the performance of these circuits is limited by decoherence caused by resonant two-level system (TLS) defects. These appear in dielectrics, such as in the alumina barrier of a Josephson junction [3,4], interlayer dielectrics [4,5], and in native oxides found on various substrate and superconductor surfaces [6,7]. As a result, the development of quantum integrated circuit technology depends on reducing these parasitic defects. In addition, because many types of defects are expected in integrated circuits, it is essential to identify the two-level systems to avoid exhaustive heuristic searches for new processes and materials.

In amorphous materials, including dielectrics, the conventional TLS model describes observed low-temperature behavior in terms of a distribution of tunneling defects [8-10]. Despite later refinements to the theory [11-13], the specific atoms or groups of atoms which tunnel are generally unknown. Arguably the best characterized amorphous dielectric is $SiO_2$, where TLS phenomena are well established by specific heat [14] and dielectric measurements [15]. In electric spin echo measurements, a TLS dipole moment from the GHz regime was found to be correlated with OH-concentration [16]. Room temperature far infrared absorption spectra, measured above the frequency of the known free OH rotor absorption, interpreted the OH motion as partial rotations around a central SiO bond due to configurations and tunneling in the solid environment [17]. By using a double well potential model Phillips *et al.* concluded that the same potential would produce a 3.7 GHz frequency difference in the lowest two energy states [18], establishing a widely accepted physical model of a TLS in the GHz regime.

Of particular interest to superconducting qubits is alumina, including amorphous $Al_2O_3$. In the majority of superconducting qubits this material is used as the Josephson junction tunnel barrier, as the substrate material, and also appears as the native oxide of the aluminum wiring [3,4,7,13]. In the alumina Josephson junction barrier, TLS are found to be consistent with the tunneling motion of OH, which are comparable to the OH-related TLS in $SiO_2$ [5,19]. In addition, dielectric relaxation measurements of $Al_2O_3$ films show that TLS density increases monotonically with $H_2$ exposure [20]. Although hydrogen is clearly indicated as a TLS defect in qubits, acting as an OH rotor or a more complicated structure, a full physical model of a TLS is still lacking.

To provide a more complete description of TLSs, we have calculated and identified bulk and surface hydrogen defects of c-$Al_2O_3$ using *ab initio* techniques. Here, we discuss bulk hydrogenated Al vacancies and interstitial hydrogen coordinated to six adjacent O atoms in the $Al_2O_3$ crystal. We also consider a surface OH rotor, where the O is attached to a surface Al atom on an $Al_2O_3$ crystal to form a solid surface analog of the canonical OH rotor. We calculate the formation energies, tunneling energies and dipole moments of the defects. Furthermore we predict the significance of these defects as TLS to devices in the GHz and THz regimes.

We first identified defects of sufficient concentration to contribute to TLS loss by calculating their formation energies ($E_f$) derived from total energy calculations [21]. Electronic structure calculations were performed using a screened hybrid density functional theory (HSE06) [22] with the amount of exact exchange adjusted to 32% (see Supplementary Material, SM [23]). The exchange adjustment did not significantly impact the predicted lattice parameters, and the calculated properties are in close agreement with experiment, as summarized in Table **1**.

**Table 1**: Calculated lattice parameters, direct band gap and formation enthalpy of α-$Al_2O_3$ with experimental values listed for comparison.

| α-$Al_2O_3$ | HSE06(32%HF) | Expt |
|---|---|---|
| a (Å) | 4.73 | 4.76[a] |
| c (Å) | 12.96 | 12.98[a] |
| Bandgap (eV) | 8.88 | 8.80[b] |
| $\Delta H_f$ (eV/f.u) | -16.39 | -17.36[c] |

[a] ref. [24], [b] ref. [25], [c] ref. [26]

We plot the predicted defect formation energies for the lowest energy charge state as a function of Fermi level and under both O-rich and Al-rich growth conditions in Fig. **1** for bulk and surface defects of α-Al$_2$O$_3$ (see SM [23]). Surface hydroxylation $[OH_{surf}]$ is found to be favorable under all growth conditions, suggesting that hydroxide termination at interfaces and of surfaces of Al$_2$O$_3$ will occur when exposed to water [27]. The resulting surface OH rotor is found to have three degenerate local minima along the rotor path. The formation of a bulk Al vacancy $[V_{Al}]^q$ results in the formation of six nearest neighbor oxygen dangling bonds near the valence band maximum. For neutral Al vacancies ($q = 0$) the O dangling bonds are occupied by three holes which become populated with increasing Fermi level (E$_F$), creating the charge states $q = -1, -2,$ and $-3$, the most energetically favorable $[V_{Al}]^q$ defect. Hydrogenation of the bulk Al vacancy by H$^+$ $[V_{Al}-H^+]^q$ creates a stable defect with charge ranging from $q = +1$ at the valence band maximum to $q = -2$, the most favorable charge across the widest range of E$_F$. The charge state of nearest neighbor O atoms becomes more negative with increasing E$_F$, leading to tighter binding with the H$^+$ defect. This results in differing structural relaxations with changing charge state, and indicates that the hydrogenated defect will have properties dependent on the charge state. Although c-Al$_2$O$_3$ is considered to be a low loss dielectric material [28], under O-rich conditions the formation of hydrogenated Al vacancy defects is found to be energetically favorable and our results suggest that these defect types should be common in amorphous Al$_2$O$_3$.

We predict interstitial hydrogen defects $[H_{int}]^q$ to occur in significant concentration in Al$_2$O$_3$ at low E$_F$ for both O-rich and Al-rich growth conditions. The stable form of interstitial hydrogen under these conditions is H$^+$, where the H$^+$ is localized to one of its six nearest neighbor oxygen atoms. As E$_F$ increases the H$^+$ defect changes charge state to an H$^-$; the neutral charge state H is never the most stable form of this defect. Upon formation of H$^-$, structural relaxation causes H$^-$ to no longer bond to a nearest neighbor oxygen but to instead occupy a defect site equidistant from its six nearest neighbor O atoms. In contrast, interstitial H$^+$ has six degenerate local minima, each localized at one of its six nearest neighbor O atoms. The adjacent minima for interstitial H$^-$ are much further apart than the minima of H$^+$, indicating that interstitial H$^+$ defects are significantly more likely to tunnel than interstitial H$^-$ defects in Al$_2$O$_3$. Interstitial molecular H$_2$ was not found to be energetically viable under any growth conditions.

The conventional TLS model in amorphous solids is based on atoms tunneling between two neighboring potential wells [8-10]. Although higher symmetry multi-well potentials exist in the crystalline form, local strain in amorphous materials is expected to distort these potentials to the double-well form. In this model, the number density distribution $d^2n = P_0\, d\Delta\, d\Delta_0/\Delta_0$ depends on the defect's tunneling energy $\Delta_0$, asymmetry energy $\Delta$ and material constant $P_0$. In this model the TLS energy $E_{TLS} = \sqrt{\Delta_0^2 + \Delta^2}$ is larger than the tunneling energy due to contribution from the asymmetry energy. However, the resonant field loss from a TLS is proportional to $\Delta_0^2/E_{TLS}^2$, such that the TLS coupling to fields is largest when asymmetry is the smallest [17]. Therefore, we expect that the center of the measured broad TLS energy distribution produced by an amorphous solid defect will be approximately equal to the tunneling energy $\Delta_0$ calculated for the same defect in the corresponding crystal.

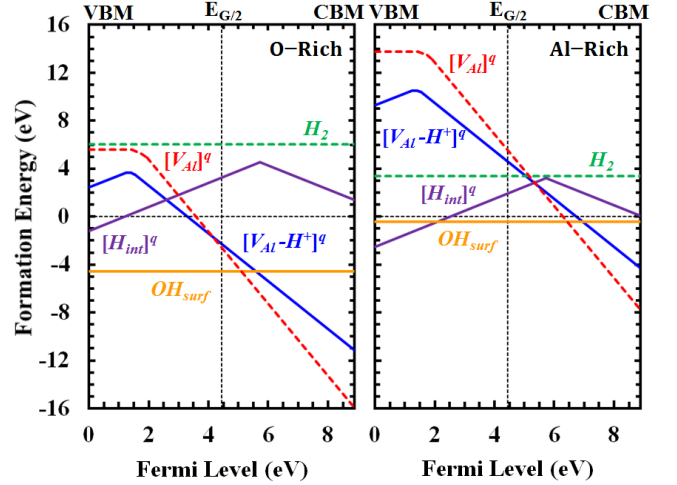

**Fig 1**: Defect formation energies (E$_f$) under O-Rich (left) and Al-Rich (right) growth conditions for interstitial H (purple), surface OH (orange), Al vacancy (dashed-red), and hydrogenated Al vacancy (blue) defects in α-Al$_2$O$_3$ as a function of Fermi level. Only the lowest energy charge state is shown for each defect type within the band gap for α-Al$_2$O$_3$. Interstitial H$_2$ (green) provided as a reference.

We determine defect tunneling from the structurally relaxed minimum energy pathway (MEP) between defect potential minima using the nudged elastic band procedure [29] for each charge state of the viable hydrogen based defects. The inherent C$_3$ symmetry axis perpendicular to the (0001) face and local S$_6$ symmetry points in α-Al$_2$O$_3$ suggest that defects formed in this material may exhibit 3-fold or 6-fold symmetric rotational character about this axis corresponding to the migration between degenerate localized defect sites. We found that surface hydroxides (OH), hydrogenated Al vacancies and interstitial H all followed a MEP corresponding to a quantum rotor (Fig. **2** top). We next determined the defect tunneling energy by fitting the calculated MEP to a rotational Hamiltonian involving $j$ rotational minima (Fig. **2** bottom),

$$H = -\frac{\hbar^2}{2I}\frac{\partial^2}{\partial\theta^2} + V_0 \cos(j\theta).$$

For $j = 2$, this equation is an exactly solvable Mathieu equation with eigenfunctions of the form of symmetric and antisymmetric combinations of the single-well ground state wavefunctions (localized basis), and are denoted as $E_{0k}$, where $k = 0$ or 1 to indicate the symmetric ground state or antisymmetric first excited state solutions. The transmission of the eigenfunctions through the barrier induces an energy splitting $\Delta_0$ between the 0 and 1 levels of the quantum rotor TLS.

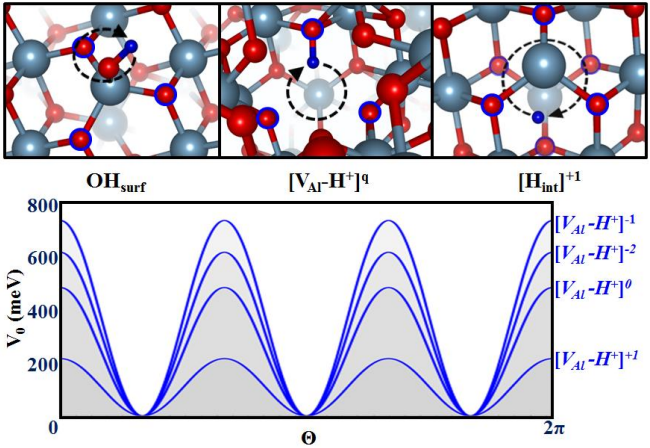

**Fig 2**: (**top**) H (blue) based tunneling rotor defects identified in $Al_2O_3$ viewed partially off axis in the [0001] direction. Dashed circles indicate rotor MEP. Blue encircled O atoms (red) indicate the local rotor minima. (left) OH rotor bound to a surface Al (gray) resulting in a 3-fold degenerate rotor. (middle) Hydrogenated Al bulk vacancy defect resulting in the formation of a 3-fold degenerate $H^+$ rotor. (right) Interstitial $H^+$ with six O nearest neighbors that form a 6-fold degenerate rotor. (**bottom**) 3-fold degenerate relaxed PES of hydrogenated charged Al vacancies in $Al_2O_3$. PES calculated along the MEP for transitions of H between minima and fit to an analytic function to solve for the TLS properties (Table **2**).

We extended the application of the Mathieu equation to model 3-fold and 6-fold symmetric rotors by a linear transform, $j\theta \rightarrow 2\theta'$. This approach only identifies the fully symmetric and fully antisymmetric eigenfunctions for both the 3-fold and 6-fold degenerate rotor potentials. A 3-fold degenerate rotor potential has three solutions in the localized basis, the $k = 0$ ground state and the $k = \pm 1$ doubly degenerate first excited state solutions, where only the $k = -1$ state is not determined from solution of the Mathieu equation. Thus, this approach provides a valid approximation for determining $\Delta_0$ of a 3-fold degenerate rotor [30].

An additional approximation is applied to solve for $\Delta_0$ of a 6-fold degenerate rotor potential, in which two pairs of degenerate levels exist among the set of six solutions in the localized basis, $k = 0, \pm 1, \pm 2,$ and 3. Here, mapping the 6-fold potential to the form of a Mathieu equation identifies the 0 and 3 eigenfunctions. Therefore, our reported $\Delta_0$ for these defects is approximated by treating the $\pm 1$ and $\pm 2$ solutions as being evenly spaced between $E_{00}$ and $E_{03}$ so that $\Delta_0 = (E_{03} - E_{00})/3$. Furthermore, the higher order degeneracy of the rotors may be reduced to that of a double well potential by additional environmental disorder due to strain or interactions with other defects. However, even for large shifts in the asymmetry energy $\Delta$ only a small perturbation to the potential barrier $V_0$ occurs, suggesting that the double-well tunneling energies would be lower by a factor on the order of unity from our calculated values.

**Table 2**: Calculated properties for each tunneling H defect type and charge state in $Al_2O_3$ identified in this study. The rotor radius (R) and potential barriers ($V_0$) are computed from the structurally relaxed rotor MEP. Dipole moments (p) are derived from the displacement between rotor potential well minima and a corresponding charge analysis. Calculated tunneling splitting energies ($\Delta_0$) for identical hydrogen (H) and deuterium (D) substituted defects are reported in GHz.

| Defect | R(pm) | p(D) | $V_0$(meV) | $\frac{\Delta_{0,H}}{h}$ (GHz) | $\frac{\Delta_{0,D}}{h}$ (GHz) |
|---|---|---|---|---|---|
| $OH_{surf}$ | 76 | 3.2 | 19.1 | $1.0 \times 10^3$ | $2.8 \times 10^2$ |
| $[V_{Al}-H^+]^{+1}$ | 71 | 3.0 | 214 | $1.3 \times 10^1$ | $2.1 \times 10^{-1}$ |
| $[V_{Al}-H^+]^0$ | 71 | 3.0 | 480 | $2.1 \times 10^{-1}$ | $4.9 \times 10^{-4}$ |
| $[V_{Al}-H^+]^{-1}$ | 68 | 2.8 | 731 | $2.2 \times 10^{-2}$ | $1.7 \times 10^{-5}$ |
| $[V_{Al}-H^+]^{-2}$ | 67 | 2.8 | 612 | $1.0 \times 10^{-1}$ | $1.5 \times 10^{-4}$ |
| $[H_{int}]^{+1}$ | 95 | 2.3 | 152 | [a] $2.4 \times 10^2$ | [a] $2.3 \times 10^1$ |
| $[H_{int}]^0$ | - | - | [b] 1300 | - | - |
| $[H_{int}]^{-1}$ | - | - | [b] 1420 | - | - |

[a] Forms a 6-fold degenerate rotor defect (*vide infra*)
[b] Does not form a rotor defect

Table 2 shows our calculated H-based defect rotor properties. Bader charge analysis [31,32] was performed to identify the charge on the tunneling atoms along the MEP. The dipole moment p was determined using the tunneling distance between local minima and the H atom charge. The surface dipole moment was found to be the largest at 3.2 Debye. The moment of the bulk defects is only slightly smaller such that they could all couple strongly to Josephson junction qubits. For the hydrogenated cation defect the moment depends on charge such that the low and high charge states should be distinguishable.

OH surface defects are found to form a 3-fold degenerate tunneling rotor with a tunneling barrier of only 19.1 meV between local minima. This leads to an energy splitting in the THz regime due to tunneling between the ground and first excited states of the rotor. However, the favorable defect formation energy for hydroxylation of the $Al_2O_3$ surface and lack of dielectric screening between the rotor and its environment suggest that the potential energy surface of the rotor is readily perturbed by defect interactions [33]. Thus, it is expected that surface OH rotors will create GHz loss in superconducting devices.

Hydrogenated Al vacancies are predicted to exist in significant concentration and are found to form 3-fold degenerate rotors that contribute to TLS loss throughout the GHz regime. A distribution of barrier heights due to the different possible defect charge states is predicted to range from 214 to 731 meV and corresponds to a TLS frequency range of 0.02 to 13 GHz. However, modulation of the Fermi level may provide a route for tuning the loss properties due to hydrogenated Al vacancies by changing the defect charge state and thus the TLS energy. Additionally, substitution of H with deuterium may offer another route for altering the loss properties due to hydrogenated defects by reducing their tunneling energies.

From the computed range of viable formation energies shown in Fig. **1** for a q=-2 hydrogenated Al vacancy, a formation energy of 0.5 eV is found to be reasonable for a slightly O-rich environment and a mid-gap Fermi energy. For an equilibrium temperature of 300C, this can easily lead to a defect concentration of $10^{18}$/cm$^3$. With the TLS asymmetry energy spread over a 100 MHz band and tunneling energy spread over a few orders of magnitude we find $P_0 \sim 3 \times 10^{45}/(Jm^3)$, where this result is only logarithmically sensitive to the range of tunneling energies. For the calculated dipole of 3 Debye, this results in a loss tangent on the order of $\tan \delta \sim 2 \times 10^{-3}$, a value consistent with loss from large-area alumina Josephson junction barriers in qubits [5].

Interstitial H is found to form a 6-fold degenerate rotor in the $q = +1$ charge state. The higher order degeneracy of this tunneling rotor leads to significant coupling between local minima and produces a TLS in the mid GHz energy range. Although these defects are only likely to occur at low Fermi levels and may be difficult to realize experimentally in c-Al$_2$O$_3$, they are predicted to create loss onset in the mid GHz regime and into the THz regime. The reduced forms of interstitial H defects, $q = 0$ and -1, do not create a rotor defect because the H is constrained to a single local minima equidistant from the six neighboring O atoms. For these charge states the nearest tunneling site is an equivalent defect site on a different set of neighboring O atoms with a barrier to tunneling > 1eV. Thus, the $q = 0$ and -1 charge states of interstitial H defects are predicted to not contribute to TLS loss at low temperatures.

The formation energy and low-temperature tunneling energies were calculated for bulk and surface hydrogen defects using *ab initio* methods. All defects were studied in alumina, a common dielectric in superconducting devices, and had a sufficient dipole moment to cause strong coupling to a superconducting qubit. Negatively-charged hydrogenated cation vacancies are found to likely form and have tunneling energies which are sufficiently low in frequency to cause loss throughout the GHz frequency range. This allows us to theoretically predict an important low temperature TLS for superconducting qubits, with a precise definition appropriate for amorphous and crystalline alumina. We also found that the interstitial H defect may cause a substantial loss with onset in the mid GHz regime and into the THz regime. Hydrogenated cation vacancy defects should create TLS in other materials, as should various other defects which can be predicted by similar methods.

This work was funded by IARPA through the U.S. Army Research Office (award No. W911NF-09-1-0351) and utilized the Janus supercomputer, which is supported by the National Science Foundation (award number CNS-0821794) and the University of Colorado Boulder.